\documentclass[AMA,LATO1COL,sort&compress]{WileyNJD-v2} %连续引用参考文献

\usepackage{lineno,hyperref}
\usepackage{multirow}
\usepackage{mathrsfs}
\usepackage{graphicx,epsfig,float}
\usepackage{algpseudocode}
\usepackage{amsmath}
\usepackage{graphics}
\usepackage{epsfig}
\usepackage{amssymb}
\usepackage{algorithm}
\usepackage{algorithmicx}
\usepackage{algpseudocode}
\usepackage{subfigure}
\hypersetup{hidelinks} %去除绿框

\articletype{RESEARCH ARTICLE}%

\received{26 April 216}
\revised{6 June 2016}
\accepted{6 June 2016}

\begin{document}

\title{PSSPR: A Source Location Privacy Protection Scheme Based on Sector Phantom Routing in WSNs}

%\author[myfirstaddress]{}
%\ead{ylwangqfnu@163.com}
%\author[myfirstaddress]{}
%\ead{yangguoyu1020@163.com}
%\author[myfothaddress]{}
%\ead{lifengzhang@ldu.edu.cn}
%\author[myfirstaddress]{}
%\ead{darcy\_wang@163.com}
%\author[mysecondaryaddress]{}
%\ead{kelishan@gzhu.edu.cn}
%\author[myfifthaddress]{}
%\ead{douyi12@gmail.com}
%\author[seventhaddress]{Shouzhe Li}
%\ead{sli649@wisc.edu}
%\author[mysixthaddress]{Xiaomei Yu}
%\ead{yxm0708@126.com}

\author[1,4]{Yuling Chen$^*$}
\author[1]{Jing Sun}
\author[2]{Yixian Yang}
\author[1]{Tao Li}
\author[2]{Xinxin Niu}
\author[3]{Huiyu Zhou}

\authormark{Yuling Chen\textsc{et al}}

\address[1]{\orgdiv{State Key Laboratory of Public Big Data}, \orgname{College of Computer Science and Technology}, \orgname{Guizhou University},\orgaddress{\state{Guiyang}, \country{China}}}
\address[2]{\orgdiv{School of Cyberspace Security}, \orgname{Beijing University of Posts and Telecommnuications}, \orgaddress{\state{Beijing}, \country{China}}}
\address[3]{\orgdiv{School of Informatics}, \orgname{University of Leicester}, \orgaddress{\state{England}, \country{United Kingdom}}}
\address[4]{\orgdiv{Guangxi Key Laboratory of Cryptography and Information Security } }

\corres{Yuling Chen, State Key Laboratory of
	Public Big Data, College of Computer
	Science and Technology, Guizhou
	University, Guiyang 550025, China.
	Email: ylchen3@gzu.edu.cn}

\abstract[Abstract]{ Source location privacy (SLP) protection is an emerging research topic in wireless sensor networks(WSNs). Because the source location represents the valuable information of the target being monitored and tracked, it is of great practical significance to achieve a high degree of privacy of the source location. Although many studies based on phantom nodes have alleviates the protection of source location privacy to some extent. It is urgent to solve the problems such as complicate the ac path between nodes, improve the centralized distribution of Phantom nodes near the source nodes and reduce the network communication overhead. In this paper, PSSPR routing is proposed as a visable approach to address SLP issues. We use the coordinates of the center node $V$ to divide sector-domain, which act as important role in generating a new phantom nodes. The phantom nodes perform specified routing policies to ensure that they can choose various locations. In addition, the directed random route can ensure that data packets avoid the visible range when they move to the sink node hop by hop. Thus, the source location is protected. Theoretical analysis and simulation experiments show that this protocol achieves higher security of source node location with less communication overhead.
}


\maketitle

\section{Introduction}
The Internet of Things(IoT) refers to real time collection of any objects or processes that need to be monitored, connected, and interacted through various information sensors. As a result of the rapid development of Internet of Things (IoT), wireless sensor networks (WSNs)\cite{ref1,ref2,ref3,ref4} are vital components of the IoT has been applied in various domains. Unlike wired networks, WSNs are flexible to adapt to complex application scenarios, such as target detection\cite{ref5,ref6,ref7},  military defense\cite{ref8,ref9},medical assistance\cite{ref10,ref11}, etc.
Sensor nodes in WSNs communicate with each other via wireless devices. These sensor nodes are used to sense and collect valuable information in the network and forward it to the sink node via multiple hops. Finally, the sink node stores and processes the collected information. However, due to the unattended and openness of WSNs, anyone with a relevant wireless receiver can detect and intercept messages among sensor nodes. The adversary may apply illegal means to communicate with powerful workstations or information sources and potential security issues will inevitably occur. Therefore, the security of the network has become a critical issue for the WSNs. Considering the practical significance of the source location, we focus on the SLP protection in this paper.

\subsection{Related works} %二级标题 相关工作

 Recently, there has been an increased focus on location privacy and many strategies attracted attentions as approachs to prevent adversaries from performing a backtracking strategy to obtain the source location. Ozturk et al. \cite {ref12} first introduces his concept. Location privacy as a contextual-oriented problem has been widely applied in various domains, such as blockchain technology\cite {ref13,ref14}, smart cities, Intelligent technology and so on. Location privacy covers the source location privacy and the sink location privacy. This paper focus on the issue of Source location privacy (SLP)protection. Spachos et al. \cite{ref15} proposes a dynamic routing scheme (ADRS) based on the Angle. However, in this scheme, each hop can only randomly select the relay node to transmit packets within the fixed Angle range, which limits the protection of the source position to a certain extent. For the above reason, Liu Ya et al.\cite {ref16} utilized variable Angle and proposes a dynamic routing scheme (VADRS) to protect the source location privacy, which improves the safety performance of ADRS by selecting the optimal Angle for data transmission at each hop. Zhang Jiang nan et al.\cite {ref17} proposes a single virtual loop routing SVCRM protocol based on fake packets, and optimized it to put forward MVCRM protocol, which complicated the adversary's tracking path and extended the security time. 

Considering a more powerful adversary, Wang et al.\cite {ref18} proposes an angle-based source location privacy protection protocol (PRLA), in which the concept of visible area was first proposed and defined a path through the visible area during phantom routing, called a failure path. Credit routing\cite {ref19} and RAPFPR protocol\cite {ref20} both ensure the source location privacy, but they do not consider the influence of visible area. The PRLA protocol calculates the forwarding probability by the offset angle of the sensor nodes to reduce the possibility of routing through the visible area. In view of the shortcomings of PRLA scheme, Chen Juan et al. \cite {ref21} proposes two algorithms using phantom nodes to protect the source location privacy. By using limited flooding, the nodes away from the source were selected as phantom nodes,which significantly increased the location diversity of phantom nodes in Source Location Privacy Preservation Protocol in Wireless Sensor Network Using Source‐Based Restricted Flooding (PUSBRF). Compared with the PUSBRF protocol, EPUSBRF avoids the visible area in the routing process and extends the network security time. Kong Xiangxue et al. proposes a source location privacy protection routing protocol based on random virtual ring(PRVR)\cite {ref22}, which extended the routing path to the virtualization range of the source node, making it difficult for adversaries to implement efficient reverse tracking. Unlike the  active attacks\cite {ref23,ref24,ref25}, the adversaries in this scheme can only launch local passive attacks.

In addition,  Kamat et al.\cite {ref26}and Kang et al.\cite {ref27} use directed random walk to protect the source location privacy. In their scheme, The phantom sources are far from the source node. According to the hops from a certain node to Sink, its neighbor nodes are classified into child and parent node sets respectively. When the forwarding node transmits the packets, it randomly selects a receiving node from the parent or child node set to send the packet to the Sink node. However, the phantom nodes obtained by the directed routing method will gather in some fixed areas, which cannot achieve the purpose of geographical diversity of the phantom nodes and complex transmission routes.

\subsection{Motivation and contributions} %二级标题 贡献

When sensor nodes are communicating with each other, with an appropriate wireless devices, a person can monitor the communication signals between pairs of nodes in the wireless sensor network. In spite of encryption techniques that protect the communication content exchanged between two sensor nodes, the adversaries mostly use powerful equipment or illegal means for locating the information sources. Therefore, many researchers have focused on source location privacy (SLP) protection in recent years. What’s more, in the process of routing packets how to avoid visible area, prolong the safe time and reduce the adversary’s detection probability are the current research focus. As a result, this paper proposes a source location privacy protection policy based on sector phantom routing in WSNs. In addition, the WSNs consist of many low-power wireless sensor nodes, it is a key consideration to improve the location security of source nodes while taking into account the performance of network nodes. 

In this paper, we focus on SLP protection and propose a source location privacy protection scheme based on sector domain phantom routing in WSNs(PSSPR), which has a good performance in security. The main contributions of this paper are as follows:
\begin{itemize}
\item We propose a phantom routing based on sector domain. We use the coordinates of the central node V to select phantom nodes and determine the candidate phantom node area, which address defects of insufficient diversity of phantom node positions. And through directed routing to ensure that the selected phantom nodes are distributed in an area far from the source node. 

\item Due to the generated phantom node is located between the source node and the base station, our proused scheme uses a smaller communication overhead to achieve higher source location security.

\item This scheme avoids the visible area and extends the path length of the adversary's backtracking data packet.
\end{itemize}

\subsection{Roadmap} %二级标题 

The remainder of this paper is organized as follows. The network model and the advversary models are presented in section 2. Section 3 gives an overview of our proposed scheme. Section 4 evaluates the performance of our proposed scheme,whilst the security analysis is outlined in section 5. Finally,we conclude the paper and the describe planned future studies in section 6. 

%第二章
 \section{PRELIMINARIES}  %一级标题 相关知识

\subsection{Network model} %二级标题 网络模型

In this paper, a typical Panda-Hunter model is used to study SLP protection. As shown in Fig.1. Pandas are assumed to inhabit a monitoring area with a large number of randomly and uniformly distributed sensor nodes. Once detecting a panda, the sensor node becomes the source and regularly reports the monitored message to the sink node in a multi-hop manner. The hunter can acquire the Immediate sender node's location in the way of backtrace packets by analyzing the transmission signal. The operation is repeated until the hunter reaches the source node and stops. When the hunter enters the visible area of the source location, the source location exposed. For the network, we make the following assumptions: 1) The sink node is regarded as the final destination of all data packets and it remains in the entire network center.The location of all sensor nodes remain unchange after deployment. 2) Any two sensor nodes in the network communicate through one or multi-hops mode. 3) All sensor nodes in the network are randomly and uniformly deployed in the sensored area. In addition, All sensor nodes have the same characteristics, which means that they have the same computing ability, initial energy, and cache memory. 

%图1
\begin{figure}[htbp]
	
	\centering
	
	\includegraphics[height=6.0cm,width=9.5cm]{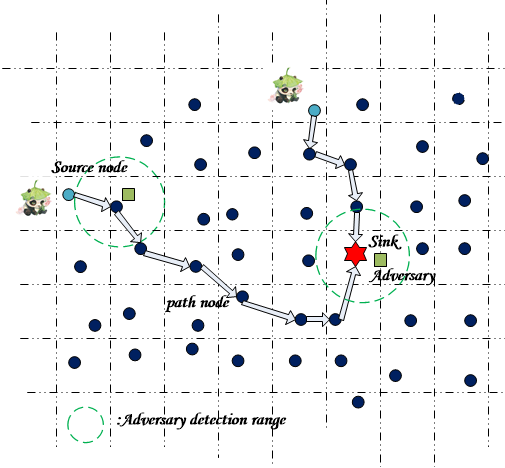}%fig2文件夹下的xbee.esp图片，
	
	\caption{The Panda-Hunter model}
	
\end{figure}

\subsection{Attack model} %二级标题 攻击者模型
In a typical $SLP$ protection scenario, we assume a patient attacker mode with a small visibility of the network. The attacker can only eavesdrop any message for packets in sensor networks and cannot interfere with the normal operations of the sensors network. The data monitored by the source is regularly transmitted to the sink destination node according to a certain routing protocol. Therefore, under normal circumstances, if the attacker initially locates at the sink node, it can be guaranteed to successfully capture all packets. Driven by huge profits, the adversary is an illegal attacker that is equipped with powerful devices to eavesdrop any message from the source. When the attacker overhears a new message, it will measure the angle of arrival of the signal and the received signal strength to identify the immediate sender node, it will not have any functional impact on the network. Then, the adversary performs passive attack by starting back tracing the packet route by moving to the previous node towards the source until it reaches the source node. Algorithm 1 explains the strategy of the adversary.

\subsection{Security assumption} %二级标题 安全假设
In the entire network, we assumes that the sink node is absolutely secure. The communication between nodes adopts secure encryption and the adversary cannot obtain the location of the source through destorying the sink or intercepting data packets.

%第三章
\section{PSSPR} %一级标题
Once the assets is detected, the nodes around the area will become source node and continuously monitors assets activities and locations. The source periodically generates sensored data packets and forwards them to the sink node via multi-hops routing manner. Before sending the packet, the source will first judge whether the distance ${D_{S,B}}$ between itself and the sink is within the communication radius $ r $. If ${D_{S,B}}$ is less than or equal to the $ r $, the source will directly sends the received packet to the sink. Otherwise, the source node transmitts packets using the routing method we proposed from the current node to sink node. In this section, we will introduce details of our proposed $ PSSPR $ scheme. The proposed PSSPR is based on our noticed source location privacy based on sector domain routing. It consists of four phases: network initialization, phantom routing based on sector domain, same-hop routing and variable Angle routing. An overview of the PSSPR scheme is shown in Fig.2. 

\subsection{Network initialization} %二级标题 网络初始化

After the deployment of the wireless sensor nodes is completed, all sensor nodes have their own unique identity identification information ${ID}$ of the whole network. In the network initialization phase, the sink establishes a horizontal Cartesian coordinate system with itself as the center and then broadcast a \textit{Sink-Msg} $ \left\{ID, HopCount, Infor\right\}$ message to all sensor nodes in flood mode. \textit{ID} denotes sending node and $HopCount$ denotes the hop count from the sink which initial value is set to $ zero $. When a sensor node receives the \textit{Sink-Msg} $ \left\{ID, HopCount, Infor\right\}$ message, it compares the current \textit{HopCount} with the original ${HopCount}$. If the current ${HopCount}$ is smaller than the original \textit{HopCount}, the node update ${HopCount}$, otherwise, the node discards it. Each node only records the minimum ${HopCount}$. \textit{Infor} represents the coordinate information of the sensor nodes in the wireless network. If a node receives \textit{Sink-Msg} for the first time, it records the hop count,upgrades the value of $HopCount = HopCount+1$, and transmits the beacon message to its nerghbor nodes. For each node that receives \textit{Sink-Msg} information, the ${ID}$, ${Infor}$, and ${HopCount}$ of the forwarding node should be stored in its neighbor table. After the sink completes the whole network broadcast in flood mode, each node records the coordinate information of the sink, itself and its neighbor nodes record and storage the minimum ${HopCount}$ from the sink. Table 1 lists the main notations used in this study. 

%三线表 符号

%表1  
\begin{table}[htbp]
	\centering
	\caption{Summary of notations}
	\begin{tabular}{cc}
		\toprule  % 中部线
		Symbols&Meaning\\ 
		\midrule  % 中部线
		S&Source node\\
		Sink&Base station\\
		P&Phantom node\\  
		$ D_{S,B} $&The distance in hops between sink and source node\\
		$ D_{S,P} $&The distance in hops between phantom and source node\\
		
		V&Center node\\
		$ P_{i} $&Pseudo phantom node\\ 
		$ A_{\lambda i}, A_{\lambda i}'' $&Intermediate node\\
		$  A_{\lambda i}'' $ & $  A_{\lambda i}, A_{\lambda i}''$ are symmetric about node \textit{V}\\ 
		\(\ h_{m}\)&Same-hop routing hop count\\
		r&eavesdropping radius of adversary;Communication radius \\
		\(\ r_{0}\)&Radius of visible area\\
		\(\ N_{i,j}\)&The distance in hops between node i and node j\\
		\bottomrule  % 中部线
	\end{tabular}
\end{table}

\subsection{Phantom routing based on sector domain}%3.2

The location information of the phantom node \textit{P} is determined at this stage. In the network initialization phase, the source node obtains the coordinate information of itself and its list of neighbor nodes. Before forwarding the data packets, the source node first carries out $Rmax$ hops directed flood routing. Hence, each node within the $R_{max}$ hops receives the beacon message of the source ${S (S_{X},S_{Y})}$, and feedback a response message \textit{S-Mes} to the source node. The \textit{S-Mes} contains the coordinate information of each node within $R_{max}$ hops from the source. 
Steps are as follows:

%图2
\begin{figure}[htbp]
	
	\centering
	
	\includegraphics[height=6.0cm,width=9.5cm]{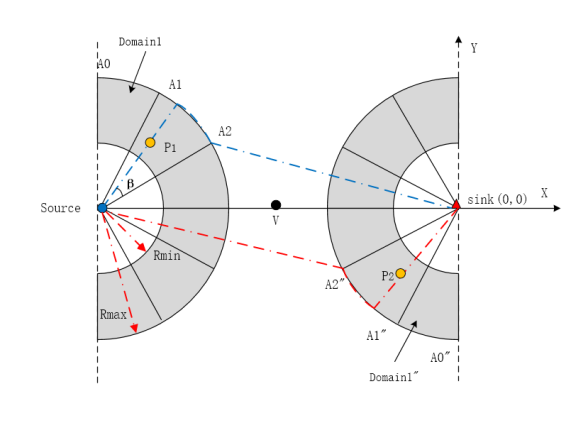}%fig2文件夹下的xbee.esp图片，
	
	\caption{Selection of phantom nodes in sector domain}
	
\end{figure}

A)  Before the source sends the data packets, it takes the sink node as the center to create a rectangular coordinate system. And the line between the source node and the sink node is the x-axis. Subsequently, the source node regards the central node between the sink node and itself as the central node $ V(V_{X},V_{Y})$.  

B)  Establish the y-axis through the convergent node and perpendicular to the x-axis. On the left side of the y axis, the source chooses a closed semicircular ring area with itself as the center, the maximum distance $Rmax$ as the outer radius, and the minimum distance $Rmin$ as the inner radius, which is called the phantom area $P_{Area1}$. The phantom area $P_{Area1}$ and the phantom area $P_{Area2}$ are symmetrical about the central node $ V(V_{X},V_{Y}) $. $P_{Area1}$ and $P_{Area2}$form the candidate domain.

C)  Divide $P_{Area1}$ into $\omega$ sector areas evenly. $\omega$ is an even number, then the angle $\theta $ of each sector area is $\frac{\pi }{\omega }$. These sector areas are designated as Domain1, Domain2, . . . , DomainW. 

D) Before transmitting packets, the source randomly selects a sector Domain$\lambda i$, $\lambda i$ $\in [1,\omega]$ and $\lambda i$ randomly distributed. The angle $\beta \in  [(\lambda i-1) \theta, \lambda i \theta]$ of the line between the phantom node and the source and the line between the source node and the intermediate node is randomly distributed
%算法1
\begin{algorithm}[t]
	\caption{Patient Adversary Algorithm} %算法的名字
	\begin{algorithmic}[1]
		\State Adversary's initial position = Sink's position ; % \State 后写一般语句
		\State When a packet is received at the sink; % \State 后写一般语句
		\State  Adversary's location = Immediate sender node's location; % \State 后写一般语句
		\While{(Adversary's location != Source's location)} % While语句，需要和EndWhile对应
		\State  Adversary's location = Immediate sender node's location;
		\EndWhile
		\State 	Adversary's location = Source's location; % \State 后写一般语句
	\end{algorithmic}
\end{algorithm}
%算法2
\begin{algorithm}[t]
	\caption{From the source to the phantom nodes} %算法的名字
	\begin{algorithmic}[1]
		\State Establish coordinate system and divide candidate domains into sectors; % \State 后写一般语句
		\State Randomly choose $ Domain_{\lambda i}( \lambda i  = 1,2 \cdots \mu )$ from candidate domains; % \State 后写一般语句
		\State 	Randomly choose $ P_{i} $ node as the phantom node P from candidate domains; % \State 后写一般语句
		\State 	Calculate angle $ \beta $; % \State 后写一般语句
		
		\If{$ P_{x} \leqslant V_{x} $} % If 语句，需要和EndIf对应
		\State send the packet through $ h $ hops $ \left (  h\in \left [ R_{min}, R_{max}\right ]  \right ) $ directed routing to reach the phantom node \textit{ P}, save $  N_{S,P}= h $;
		\State Continue directed routing away from the source until reaches $N_{S,i} = R_{max}$;
		\State The position of $ A_{\lambda i} $ is calculated from the coordinates of the phantom node $P$;
		\Else
		\State 	Rolls back to the variable Angle routing path phase;
		\State  Perform the same-hop routing phase;
		\State  Refer to the Sink-Msg, perform h-hops directed routing to the sink and determine the position of the node $P$ in this process;
		\EndIf
		
	\end{algorithmic}
\end{algorithm}

The source randomly selects a node in Domain$\lambda i$ ($\lambda i = 1,2. . .\mu $) as the pseudo-phantom node $P_{1}$ during the $kth (k \geq 1)$ packet transmission, and calculates the pseudo-phantom node $P_{2}$ according to the coordinate information of the central node $V$.  $P_{1}$ and $P_{2}$ constitute a set of pseudo-phantom nodes.

\subsection{Same-hop routing}%二级标题 3.3同跳路由

After the $R_{max}$ hops directed routing or variable included Angle routing phase is completed, the forwarding node uses the angle $\beta$ to calculate the number of same-hop routes $h_{m}$. As exhibited in formula (1):

%公式1
\begin{equation}
	h_{m}=\frac{\beta }{180^{\circ}}R_{max}
\end{equation}

If the phantom node is selected from the left side of the center node, the angle $\beta$ is calculated as:
%公式2
\begin{equation}
	\beta= \\arccos \left ( \frac{{d}_{S,A\lambda i}2\dotplus {d_{S,P}2}-{{d_{A\lambda i,P}}^{2}}}{2{d_{S,A\lambda i }}\times d{_{S,P}}} \right ) 
\end{equation}

Otherwise:
%公式3
\begin{equation}
	\beta= \\arccos \left ( \frac{{d}_{Sink,A\lambda i''}2\dotplus {d_{Sink,P}2}-{{d_{A\lambda i'',P}}^{2}}}{2{d_{Sink,A\lambda i'' }}\times d{_{Sink,P}}} \right ) 
\end{equation}

Then the forwarding node performs\(\ h_{m}\) hops in the direction close to the x-axis, and transfer the data packet to the \(\ A_{\lambda i}\) node or  \(\ A_{\lambda i}''\) node. Such node is called the intermediate node in our proposed scheme. The rules for selecting the next-hop node in the forwarding process are as follows: the phantom node $P$ selects a node  \(\ D_{S,B}\) hops away from the source or a node with the same hops as the sink in its neighbor table as the next hop forwarding node. After the \(\ h_{m}\) hop, the data packet is transmited to the\(\ A_{\lambda i}\) node or the\(\ A_{\lambda i}''\) node and then stops. This routing process is called the same hop routing. 

The same-hop routing can be analyzed by two participants. On the one hand, for an attacker, performing multiple simultaneous jumps will make the attacker mistakenly think that he is trapped in a circular trap to confuse the attacker.
On the other hand, for the source node, the same-hop routing extends the length of the attacker's reverse backtracking path and increases the security time.

\subsection{Variable angle routing path }%3.4可变角度

We use the vector inner product method to calculate the angle\(\ \varphi _{i}\).\(\ <V_{i,Mi} >\)  and \(\ <V_{S,Sink} >\)respectively represent the vector from node i to Mi and the vector from source to the sink. The calculation formula of Angle is shown in formula (4). 
%算法3
\begin{algorithm}[t]
	\caption{Variable Angle Routing Path} %算法的名字
	
	\begin{algorithmic}[1]
		\State Utilize $ <V_{i,Mi} > $ and $ <V_{S,Sink} > $ to calculate the angle $ \varphi _{i} $; % \State 后写一般语句
		\State Select the neighbor node $i$ with the smallest included Angle\(\ \varphi _{i}\) as the receiving node of the packet;
		\While{the forwarding node is common node} % While语句，需要和EndWhile对应
		\State Go back to step one;
		\EndWhile
		\If{forwarding node is the sink} % If 语句，需要和EndIf对应
		\State Stop forwarding packets;
		
		\Else
		\State End the variable routing path;
		\EndIf	
		
	\end{algorithmic}
\end{algorithm}
%公式4
\begin{equation}
	\varphi _{i}=\arccos \left [ \frac{< V_{i,Mi},V_{S,Sink}> }{\left \|V_{i,Mi}  \right \|\times \left \|V_{S,Sink}  \right \|} \right ]
\end{equation}

The forwarding node selects the node with the smallest angle \(\ \varphi _{i}\) as the next hop node. This stage is completed when the forwarding node is the sink or an intermediate node. The specific content is explained in Algorithm 3. In this stage, the forwarding direction of the next hop node is determined by the angle, so that the data packet is selected to the sink node in an approximately straight line. Under the condition that the source location is safe, this solution can control network communication consumption to a certain extent. In this stage, data packets are transmitted to sink nodes in an approximate straight line to save communication overhead  

\section{SECURITY ANALYSIS} %一级标题 第4章
\subsection{Safety analysis} %二级标题 安全性分析
In this section, we evaluate the safety of our proposed PSSPR protocol through theoretical analysis from four metrics. Random directed path, the distance from the phantom node to the source \(\ D_{S,P}\), the number of random phantom nodes $N$ and the failure path are the four influencing metrics.  Namely, by comparing the three factors with the the HBDRW scheme and the PUSBRF scheme,we come to the conclusion that the influence on safety of PSSPR is stronger. \(\ D_{S,P}\) refers to the hop count between the source node and the phantom node. If the filtered phantom nodes are clustered near the source, adversaries can capture the source node by tracing back a short routing path. Thus, the method does not protect the privacy of the source location.
And $N$ affects the diversity of the path from the source to the phantom nodes. Therefore, the greater the number of phantom nodes, the more effective the security of the source location can be protected. Eventually, The transmission path of the source forwarding packets to the sink passes through the path of the visible area, which is called the failure path. Avoiding the failure path is equivalent to extending the effective path.
\subsubsection{Random directed path} %小a
The path of a data packet from the source node to the phantom source node after a directed random h-hop is defined as a random directed path. When the attacker reaches a certain phantom source node through reverse tracking of the data packet, the source location privacy protection protocol generates a random path. The more paths there are, the more difficult it is for an adversary to trace the true source node. In a large-scale sensor network with uniformly distributed nodes, the phantom source nodes in the source location privacy protection protocol proposed in this paper are evenly distributed in a closed ring area centered on $S$, the maximum distance $R_{max}$ is the outer radius, and the minimum distance $R_{min}$ is the inner radius. Compared with the HBDRW protocol, the PUSBRF protocol increases the random directed path generated by the HBDRW protocol by $\xi$ =1-4$\gamma$/2$\pi$ = 1-2\(\ \arccos \frac{h-1}{h}\)/$\pi$.
In addition, compared with the PUSBRF protocol, the random directed path generated by the PSSBR protocol has increased: $\xi$ = 1-$h$/($R_{min}+R_{min}+1+...+R_{max}$).

%表2  
\begin{table}[htbp]
	\centering
	\caption{Percentage of random directeg path}
	\begin{tabular}{ccccc}
		\toprule  % 中部线
		h&\(\ R_{min}\)&\(\ R_{max}\)&\(HBDRW/PUSBRF\)&\(PUSBRF/PSSPR\)\\ 
		\midrule  % 中部线
		5&4& 6&40.97&33.33\\
		10&8& 12&28.71&20.00\\
		15&12& 18&23.38&14.29\\  
		20&16& 24&20.22&11.11\\
		25&22& 28&18.07&14.29\\
		30&26& 32&16.48&14.78\\ 
		\bottomrule  % 中部线
	\end{tabular}
\end{table}
With the increase of $h$, the average number of random directed paths increased by the PSSPR protocol will increase. Table 2 shows that the PSSPR protocol can significantly increase the number of random directed paths, thereby effectively improving the security of the source location privacy.

\subsubsection{Failure path}%小c

The failure path refers to the transmission path of the phantom node forwarding data packets to the sink node passes through the path of the visible area. The SLP protocol in WSN based on locational angle prove the probability of a failure path in a sensor network with uniformly distributed nodes is:

\begin{equation}
	(\arcsin( r_{0}/H )+\arcsin( r_{0}/h ))/\pi 
\end{equation}

It shows that the larger the $r_{0}$, the smaller the $H$ and $h$, the greater the probability of the transmission path passing through the visible area.In our proposed scheme, on the left side of the y axis, the source chooses a closed semicircular ring area with itself as the center, the maximum distance $Rmax$ as the outer radius, and the minimum distance $Rmin$ as the inner radius, which is called the phantom area $P_{Area1}$. The phantom area $P_{Area1}$ and the phantom area $P_{Area2}$ are symmetrical about the central node $ V(V_{X},V_{Y}) $. We know that a certain node u in the visible area, $HopCount_{u,s}\leq r_{0}$. So the phantom node selected by this method completely bypasses the visible area in the routing path of tranmitting data packets, avoiding the generation of failure path.

\subsubsection{The distance from the phantom nodes to the source\(\ D_{S,P}\)} %小a
\begin{itemize}
	\item HBDRW protocol selects the receiving node according to the hop count between the neighbor nodes of the forwarding node and the sink node. Phantom nodes are mainly distributed on a circle with $S$ as the center, radius $h$, and arc\(\ 4\gamma\)where\(\ \gamma\)is \(\ \arccos \frac{h-1}{h}\). For comparison, let\(\ h=R_{min}+h_{x}\). For the HBDRW protocol, after random $h$ hops, the average distance from the phantom nodes to the source is:
	
	%公式5
	
		\begin{equation}
			\overline{D_{S,P\ HBDRW}}=r(R_{min}+h_{x})    \left (  h\in \left [ 0,R_{min}-R_{max}\right ]  \right )
		\end{equation}

	\item Due to the source node performs $h$ hop directed random routing,the selected phantom nodes lies in the circumference with \textit{S} as the center and \(\ h=R_{min}+h_{x}\) as the radius. Correspondingly, the average distance from the phantom node to the source node in the PUSBRF protocol is: 
	
	%公式6

		\begin{equation}
			\overline{D_{S,P\ PUSBRF}}=r(R_{min}+h_{x})  \left (  h\in \left [ 0,R_{min}-R_{max}\right ]  \right )
		\end{equation}

	\item In order to achieve a high level of phantom node distribution, it is evident from Fig. 2. that The phantom nodes are distributed in the shaded area. The distance between the phantom node and the source node is between $Rmin$ and $Rmax$, so the average distance from the phantom nodes to the source is expressed as:
	
	%公式7

		\begin{equation}
			\overline{D_{S,P\ PSSPR}}=\frac{R_{min}+R_{max}}{4}\dotplus
			\int_{0}^{\frac{\pi }{2}}\frac{\sqrt{H^{2}+(R_{min}+R_{max})^{2}-2(R_{min}+R_{max})H\cos \alpha }}{\pi /4}
		\end{equation}

	As as shown in Table 3, in order to ensure the same simulation conditions of the three protocols, the corresponding values of $R_{min}$ and $R_{max}$ are set for different directed routing hop count. 

%表2  
\begin{table}[htbp]
	\centering
	\caption{Values of \(\ R_{max} \) and \(\ R_{min}\) }
	\begin{tabular}{cccc}
		\toprule  % 中部线
		h&\(\ R_{min}\)&\(\ R_{max}\)&\(\ D_{S,P\ PSSPR}\)\\ 
		\midrule  % 中部线
		5&4& 6&31.40\\
		10&8& 12&32.03\\
		15&12& 18&33.25\\  
		20&16& 24&34.64\\
		25&22& 28&36.10\\
		30&26& 32&37.42\\ 
		\bottomrule  % 中部线
	\end{tabular}
\end{table}
\end{itemize}

\subsubsection{The number of random phantom nodes N}%小b
\begin{itemize}
	\item Since the number of phantom nodes affects the random path diversity from source nodes to phantom nodes, the number of phantom nodes plays an important role in safety performance. The more phantom nodes there are, the less likely an adversary is to locate the source by backward and hop-by-hop tracking packets. 
	In PUSBRF protocol, the phantom nodes are concentrated on the circle with $S$ as the center of the circle and the radius is \(\ (R_{min}+h_{x})\) hops. And the number of phantom nodes is denotes as\(\ N_{PUSBRF}\).
	
	%公式8
	
		\begin{equation}
			N_{PUSBRF}=2\pi (R_{min}+h_{x})    \left (  h\in \left [ 0,R_{max}-R_{min}\right ]  \right )
		\end{equation}

	\item In HBDRW protocol, the phantom nodes are distributed on the arc with $ S $ as the center of the circle, a radius of \(\ (R_{min}+h_{x})\) hops, and an center angle of $ 4\gamma $ where $ \gamma $ is  ${(R_{min}+h_{x}-1)}/{(R_{min}+h_{x})}$. The number of phantom nodes is designated as\(\ N_{HBDRW}\).
	
	%公式9
	
	\begin{equation}
		N_{HBDRW}= 4\gamma\times(R_{min}+h_{x})=4\arccos\frac{R_{min}+h_{x}-1}{R_{min}+h_{x}}\times(R_{min}+h_{x})\left (  h\in \left [ 0,R_{min}-R_{max}\right ]  \right )
	\end{equation}
	
	\item In our proposed scheme, the phantom node is concentrated in the ring area with $S$ as the center of the circle and \(\ \left (  R\in \left [ 0,R_{min}-R_{max}\right ]  \right )\) as the radius. We denote the number of phantom nodes in this scheme is\(\ N_{PSSPR}\). Then,the number of phantom nodes in PSSPR scheme is approximate:
	
	%公式10
	\begin{equation}
		N_{PSSPR}=2\pi (R_{min}+h_{x})\times\ (\frac{h_{x}
			+h_{x}+1+\cdots +R_{max}-R_{min}}{h_{x}})    
	\end{equation}
	
	The number of phantom nodes produced by the three protocols is compared as shown in Table4.
	
	%表3 
	\begin{table}[htbp]
		\centering
		\caption{Comparison of the number of phantom nodes }
		\begin{tabular}{cccccc}
			\toprule  % 中部线
			h& $R_{min}$ & $R_{max}$ & $ N_{HBDRW}$&$ N_{PUSBRF}$ & $ N_{PSSPR}$\\ 
			\midrule  % 中部线
			5&4& 6&12.87& 31.42&94.24\\
			10&8& 12&18.04& 62.83&282.74\\
			15&12& 18&22.03& 94.25&565.48\\  
			20&16& 24&25.40& 125.66&942.47\\
			25&22& 28&28.38& 157.08&942.47\\
			30&26& 32&31.07& 188.50&706.86\\ 
			\bottomrule  % 中部线
		\end{tabular}
	\end{table}
\end{itemize}

\subsection{Communication overhead analysis} %二级标题 通信开销分析
In the PUSBRF strategy, the source needs to perform \(\ (R_{min}+h_{x})\) hops flooding, which is expressed as the\(\ (R_{min}+h_{x})\) hops finite flooding hop value from the source to the phantom nodes. Then the data packet is forwarded from the phantom node to the sink in the shortest route, so the average communication overhead of the PUSBRF protocol is:

%公式11
\begin{equation}
	\overline{E_{PUSBRF}}=R_{min}+h_{x}+\int_{0}^{\pi }\frac{\sqrt{H^{2}+\left (R_{min}+ h_{x}\right )^{2}-2\left ( R_{min}+ h_{x} \right )H\cos\alpha }}{\pi} d\alpha      
\end{equation}

$ \overline{E_{PUSBRF}}$ is the average communication overhead of the PUSBRF protocol; \(\ (R_{min}+h_{x})\)  is the hop length from the source node S to the phantom node P; $H$ is the hop distance between the source and the sink.
Phantom nodes generated by HBDRW are distributed on the circumference of a circle with $S$ as the center, \(\ (R_{min}+h_{x})\)  as the radius, and $ 4\gamma $ as the radian,where $ \gamma $ is  ${(R_{min}+h_{x}-1)}/{(R_{min}+h_{x})}$ then the average communication overhead of HBDRW protocol is: 

%公式12
\begin{align}
	&\overline{E_{HBDRW}}=R_{min}+h_{x}+\int_{0}^{\gamma }\frac{\sqrt{H^{2}+\left (R_{min}+ h_{x}\right )^{2}-2\left ( R_{min}+ h_{x} \right )H\cos\alpha }}{2\gamma } d\alpha + \notag\\&\int_{\pi}^{\pi+\gamma  }\frac{\sqrt{H^{2}+\left (R_{min}+ h_{x}\right )^{2}-2\left ( R_{min}+ h_{x} \right )H\cos\alpha }}{2\gamma } d\alpha     
\end{align}

For this protocol, the abscissa of the selected phantom node is smaller than the central node as an example. The data packet arrives at the phantom node after \(\ (R_{min}+h_{x})\)  hops from the source node. After the phantom node is selected, it continues to the direction away from the source node to reach$R_{max}$ , and then the same Jump, and finally in the variable angle routing path stage. Among them, the communication cost of the sector domain phantom routing is $\overline{E_{1}}=R_{max}$, the communication cost of the same hop number routing is:

%公式13
\begin{equation}
	\overline{E_{2}}=\overline{h_{m}}=\int_{0}^{\pi}\frac{\beta \pi }{180^{\circ}}R_{max}d\theta   
\end{equation}
And the communication cost of the variable angle routing path stage is:

%公式14
\begin{equation}
	\overline{E_{3}}=2[H-R_{max}+\sum_{i=1}^{\frac{\mu}{2}-1}\sqrt{H^{2}+R_{max}^{2}-2R_{max}H\cos\left (i\theta   \right ) }] /\mu   
\end{equation}

Then, the average communication overhead of PSSPR scheme is:

%公式15
\begin{equation}
	\overline{E_{PSSPR}}=R_{max}+h_{m}+2(H-R_{max}+\sum_{i=1}^{\frac{\mu}{2}-1}\sqrt{H^{2}+R_{max}^{2}-2R_{max}H\cos\left (i\theta   \right )} )/\mu     
\end{equation}

\section{SIMULATION} %一级标题 第5章
\subsection{Safety performance } %二级标题 安全时间对比分析开销
In this section, we simulate the performance of our proposed PSSPR protocol in MATLAB platform. The simulation environment and parameter configuration are as follows. There are  $ 10000 $ sensor nodes Random and evenly distributed over the monitored area of $ 6000m $ × $ 6000m $. And the communication radius  $ r $ of each node is $ 100m $ and is one third of the the visible area. The eavesdropping radius of the adversary is equal to the communication radius  $ r $. In addition, the sink node is deployed in the center of the network. In the experiment, $ \omega=6 $, and the result is shown in Fig.3.  
 
%图3
\begin{figure}[htbp]
	
	\centering
	
	\includegraphics[height=6.0cm,width=9.5cm]{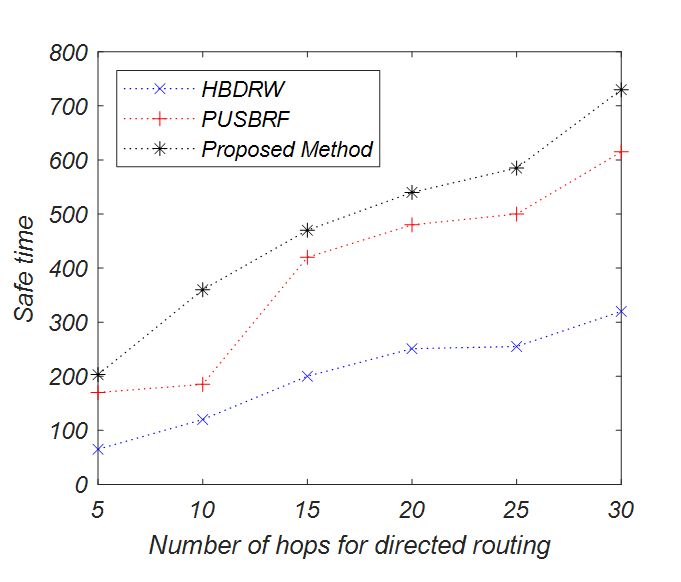}%fig2文件夹下的xbee.esp图片，
	
	\caption{Safe time VS directed routing}
	
\end{figure}

The safety time is defined as the total number of data packets sent by the source before being captured by the adversary. Fig.3 illustrates the results of $ 50 $ simulations of PSSPR scheme under the condition that hop count \textit{H} is fixed at 60. In the figure, the safe time of this protocol is extended as the directed routing $ h $ increases. It can be seen that the safety performance of PSSPR scheme is the best. Therefore, the adversary spends more time trying to capture the source node. As shown in Fig.3, the safety time of the PSSPR scheme is  $ 2.38 $  times that of the HBDRW protocol and $ 1.22 $ times that of the PUSBRF protocol. As the increases of $ h $, the average distance between the source node and the phantom node is prolonged, which increases the diversity of the directed random path and enhances the complexity of the transmission path at the same time. The complexity of the routing path makes it more difficult for the adversary to track backwards. This results in the longest safe time for the PSSPR protocol among the three protocols. 

%图4
\begin{figure}[htbp]
	
	\centering
	
	\includegraphics[height=6.0cm,width=9.5cm]{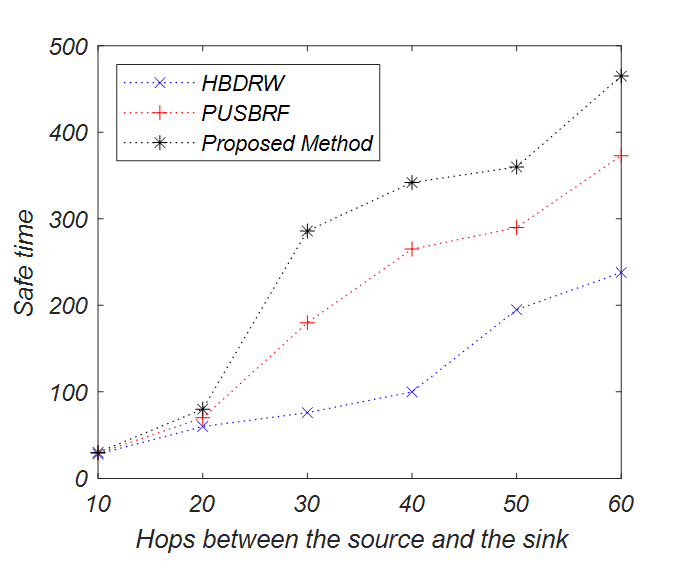}%fig2文件夹下的xbee.esp图片，
	
	\caption{Safe time VS distance to sink}
	
\end{figure}

Fig.4 displays the results of $ 50 $ simulations of PSSPR scheme under the condition that random directed routing $ h $ is $ 15 $. This result is attributed to the increase of $ H $.
As the distance to the sink node increases, the routing paths between the source node and the sink node become longer on average. In addition, the length of the transmission path increases.
Therefore, the adversary spends more time on the backtracking process due to the longer routing path. Meanwhile, our propose PSSPR protocol completely avoids the visible area in the process of transmitting packets, and thus, the safe time increases. As shown in Fig.4, the safety time of PSSPR scheme is $ 2.42 $ times of HBDRW scheme and $ 1.29 $ times of  PUSBRF scheme respectively. So our proused scheme performs clearly better than the other schemes.

\subsection{ communication overhead} %二级标题 通信开销对比分析
To balance the communication overhead and maximize the safety time, we conduct a theoretical
analysis of the communication consumption. In this experiment, the communication overhead refers to the average hops count by transmitting one data packet from the source node to the sink node.
we can analyze the simulation results of communication consumption based on four parts: the whole network flood consumption, sector-based routing path consumption, the same-hop routing consumption and the variable angle routing path consumption. Since the three protocols have the overhead of the whole network flood, our paper only considers the latter three parts. As demonstrated in Fig. 5, When the hop count $H$ between the source node and the sink node is fixed as 60, under the condition of different hops of directed routing, the average communication overhead result obtained by tranmitting 50 packets from the source node.As the random directed routing hops increases, the communication overhead of the three routing protocols become higher on average. This occurs because the larger the $h$, the longer the transmission path between the source node and the phantom node is. In addition, this stage does not greatly reduce hop count for the data packets to be routed to the sink, thus increasing the communication Overhead. It can be seen from Fig. 5 that the communication overhead of this scheme changes slightly with  directed routing hops. Fig. 5 shows that when $h$=30, the communication overhead of the three routing protocols differs the most. The communication overhead of PSSPR scheme is 7.73$\%$ and 10.04$\%$ less than that of HBDRW and PUSBRF respectively.

%图5
\begin{figure}[htbp]
	
	\centering
	
	\includegraphics[height=6.0cm,width=9.5cm]{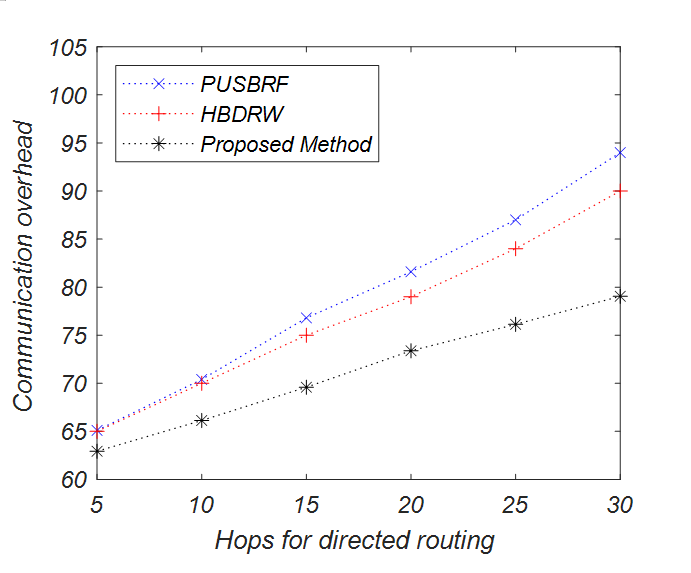}%fig2文件夹下的xbee.esp图片，
	
	\caption{Communication overhead VS directed routing hops}
	
\end{figure}

In Fig. 6, we shows the changes in the average communication overhead for different distances from the source nodes to the sink node.The simulation of the safety time is performed in terms of the hop count of random directed routing is fixed at h=20 and the source node sends 50 data packets to the sink node.It is observed that communication overhead of the three routing protocols exhibits a decreasing trend with the increasing of $H$.
This is because with the increase of $H$, more sensor nodes participate in transmitting packets, which results in an growth in transmission path. Therefore, the communication overhead increases.
We also observe that the communication overhead of the three routing protocols is not much different under the condition of the same value of $h$. From an overall point of view, our proposed PSSPR scheme contributes only slightly to communication overhead, which is 9.14$\%$ less than HBDRW and 11.57$\%$ less than PUSBRF.
%图6
\begin{figure}[htbp]
	
	\centering
	
	\includegraphics[height=6.0cm,width=9.5cm]{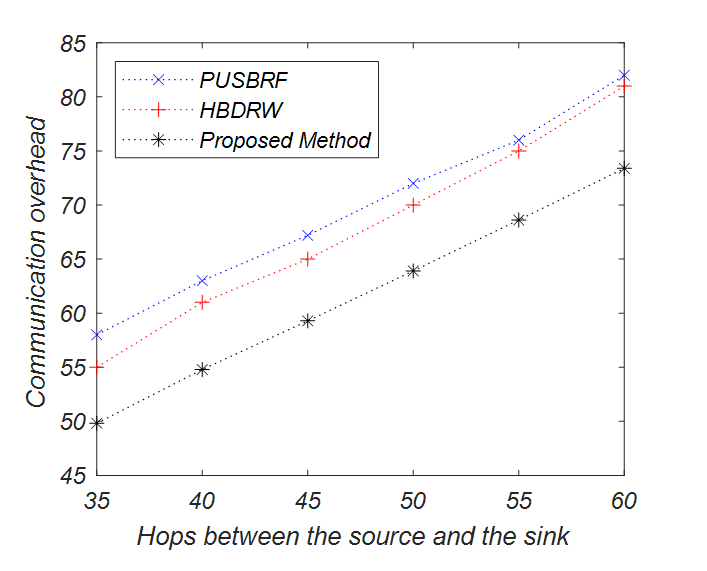}%fig2文件夹下的xbee.esp图片，
	
	\caption{Communication overhead VS distance to sink}
	
\end{figure}

\section{CONCLUSION} %一级标题 第6章
The application of wireless sensor network in the field of monitoring needs to design an effective SLP routing scheme. The existing SLP protection schemes mainly protect the SLP by changing or increasing the routing path length, which greatly increases the communication overhead. In this paper, we presented the PSSPR scheme for source location privacy, which decreases communication consumption in the interest of maintaining safety time. The scheme provides a dynamic route generation process with randomly selected phantom nodes. In the PSSPR, initially, the source node and sink node jointly establish several virtual sectors domains. Then, the phantom nodes are selected from the different sectors and the packet is transmitted through different phantom nodes to construct the dispersed routing paths to achive high security performance of the source. Along with the PSSPR protocol completely avoids the failure path and generates enormous phantom nodes while imporving the geographic diversity of the phantom nodes. Therefore,This scheme adds random directed routes, implements multiple paths and reduces overlapping paths. The overall comparative analysis shows that PSSPR protocol proves itself to be an effective scheme in the considered performance elements.

In our future work, we plan to investigate the times each node participates in routing. Then, We will comprehensively consider communication overhead and balance the energy consumption of the entire network, so that the network has a longer security time. 

\section*{Acknowledgment}%基金
This study is supported by Foundation of National Natural Science Foundation of China (Grant Number: 61962009); Major Scientific and Technological Special Project of Guizhou Province (20183001); Science and Technology Support Plan of Guizhou Province ([2020] 2Y011); Foundation of Guizhou Provincial Key Laboratory of Public Big Data (2018BDKFJJ013); Foundation of Guangxi Key Laboratory of Cryptography and Information Security(GCIS202118).

\end{document}